\documentclass[fleqn,usenatbib]{mnras}

\usepackage{newtxtext,newtxmath}

\usepackage[T1]{fontenc}

\DeclareRobustCommand{\VAN}[3]{#2}
\let\VANthebibliography\thebibliography
\def\thebibliography{\DeclareRobustCommand{\VAN}[3]{##3}\VANthebibliography}

\usepackage{graphicx}	

\usepackage{amsfonts, amsmath, amssymb, amsthm}
\usepackage{wasysym}
\usepackage{tabularx}
\usepackage{fontawesome5}
\usepackage{array}
\usepackage{subcaption}
\usepackage{float}
\usepackage{multirow}
\usepackage{setspace}
\usepackage{titlesec}
\usepackage{booktabs}
\usepackage{circuitikz}
\usepackage{hyperref}

\def\ks{km\,s$^{-1}$}

\def\cm3{cm$^{-3}$}

\def\2{$^{12}$CO}
\def\3{$^{13}$CO}
\def\8{C$^{18}$O}

\def\cm2{cm$^{-2}$}

\title[Outflows collision in EGO 338.92+0.55(b)]{Unveiling the collision between molecular outflows: observational evidence and hydrodynamic simulations}

\author[Cohen Arazi et al.]{E. Cohen Arazi,$^{1,2}$\thanks{E-mail:eitan1174@gmail.com (ECA)} 
P. F. Velázquez,$^{3,1}$\thanks{E-mail: pablo@nucleares.unam.mx (PFV)}
M. E. Ortega,$^{1}$\thanks{mortega@iafe.uba.ar (MO)}  
A. Rodríguez-González,$^{3}$
E. Alquicira-Peláez,$^{3}$\and 
S. Paron,$^{1}$
P. Rivera-Ortiz,$^{4}$
and A. Esquivel$^{3}$
\\
$^{1}$ CONICET - Universidad de Buenos Aires. Instituto de Astronom\'{\i}a y F\'{\i}sica del Espacio. Buenos Aires, Argentina\\
$^{2}$ Universidad de Buenos Aires, Facultad de Ciencias Exactas y Naturales, Departamento de Física. Buenos Aires, Argentina \\
$^{3}$Universidad Nacional Aut\'onoma de M\'exico, Instituto de Ciencias Nucleares, A.P. 70-543, 04510 Ciudad de M\'{e}xico, M\'{e}xico
\\
$^{4}$Universidad Nacional Aut\'onoma de M\'exico, Instituto de Radioastronomía y Astrofísica, Morelia, M\'{e}xico
}

\date{Accepted XXX. Received YYY; in original form ZZZ}

\pubyear{\the\year{}}

\begin{document}
\label{firstpage}
\pagerange{\pageref{firstpage}--\pageref{lastpage}}
\maketitle

\begin{abstract}
We present an unexplored scenario for interpreting the outflows in the EGO G338.92+0.55 (b) region (hereafter, EGO G338). Within this framework, we investigate the hypothesis that the interaction between two outflows is responsible for the observed morphology and kinematics of this astrophysical object. To explore this possibility, we reanalyse the region using observational molecular line data. We base our analysis on maps of moments 0, 1, and 2 of the CO emission associated with the molecular outflows. Additionally, we conduct three-dimensional hydrodynamic simulations to examine the presence or absence of a collision between two jets. From our numerical results, we produce synthetic CO images to facilitate a direct comparison with observations. The findings of this study provide compelling evidence that the observed morphology and kinematics in the EGO G338 region are the result of a likely collision between two molecular outflows.
\end{abstract}




\section{Introduction}

A ubiquitous phenomenon during star formation is the generation of molecular outflows. They are produced by the ambient molecular material that is pushed in the polar direction by jets ejected at the protostar-disk scale \citep{bally2016,bj16}. Such jets and molecular outflows are crucial because they remove the angular momentum of the protostar-accretion disk system, allowing the central object to accumulate material.

Despite numerous studies on molecular outflows (e.g.\,\citealt{maud2015, yang2018, li2020, nony2023, ortega2023}), the interaction between jets/outflows of multiple protostars remains a field that has been poorly explored both observationally and numerically. \citet{cunningham2006} suggested that outflow interactions are likely in star-forming regions. The authors point out that only the most direct collisions, where the jets collide almost head-on, may alter the global characteristics of the outflow. However, this kind of collision is rare and requires stellar densities much higher than those observed.

Understanding any kind of interaction between outflows is important for unravelling the mechanisms that fuel turbulence in molecular clouds \citep{yuan2025}, counteracting its dissipation and regulating the efficiency of star formation \citep{wang2024}.

Although there are pioneering studies on the collision between outflows and molecular clouds (e.g.\,\citealt{raga2002,choi2005}) and some suggesting possible collisions between outflows (e.g.\,\citealt{cunningham2006,beltran2012}), the first solid observational evidence of a direct interaction can be found in \citet{zapata2018}. Using Atacama Large Millimeter Array (ALMA) observations, the authors revealed a partial collision between outflows in a protostellar binary system, producing an increase in the brightness of the CO emission and in the velocity dispersion at the interaction area. Later, \citet{toledano-juarez2023} documented a collision of outflow lobes with very disparate masses, and suggested that induced star formation of the lower-mass object may have occurred due to such a collision. Recently, \citet{paron2025} suggested the disruption of a jet cavity by a molecular outflow in the region G29.862$-$0.0044, pointing out that the analysed source is a likely wide binary system of young stellar objects interacting with each other.

In this work, we present an observational and numerical study of the source catalogued as EGO 338.92$+$0.55(b) (hereafter EGO\,G338), which presents evidence of outflow collision. Given the scarcity of reported solid cases of such a phenomenon in the literature, this study can provide highly relevant results on this topic.

\section{Presentation of the source}
\begin{figure}
    \centering
    \includegraphics[width=8cm]{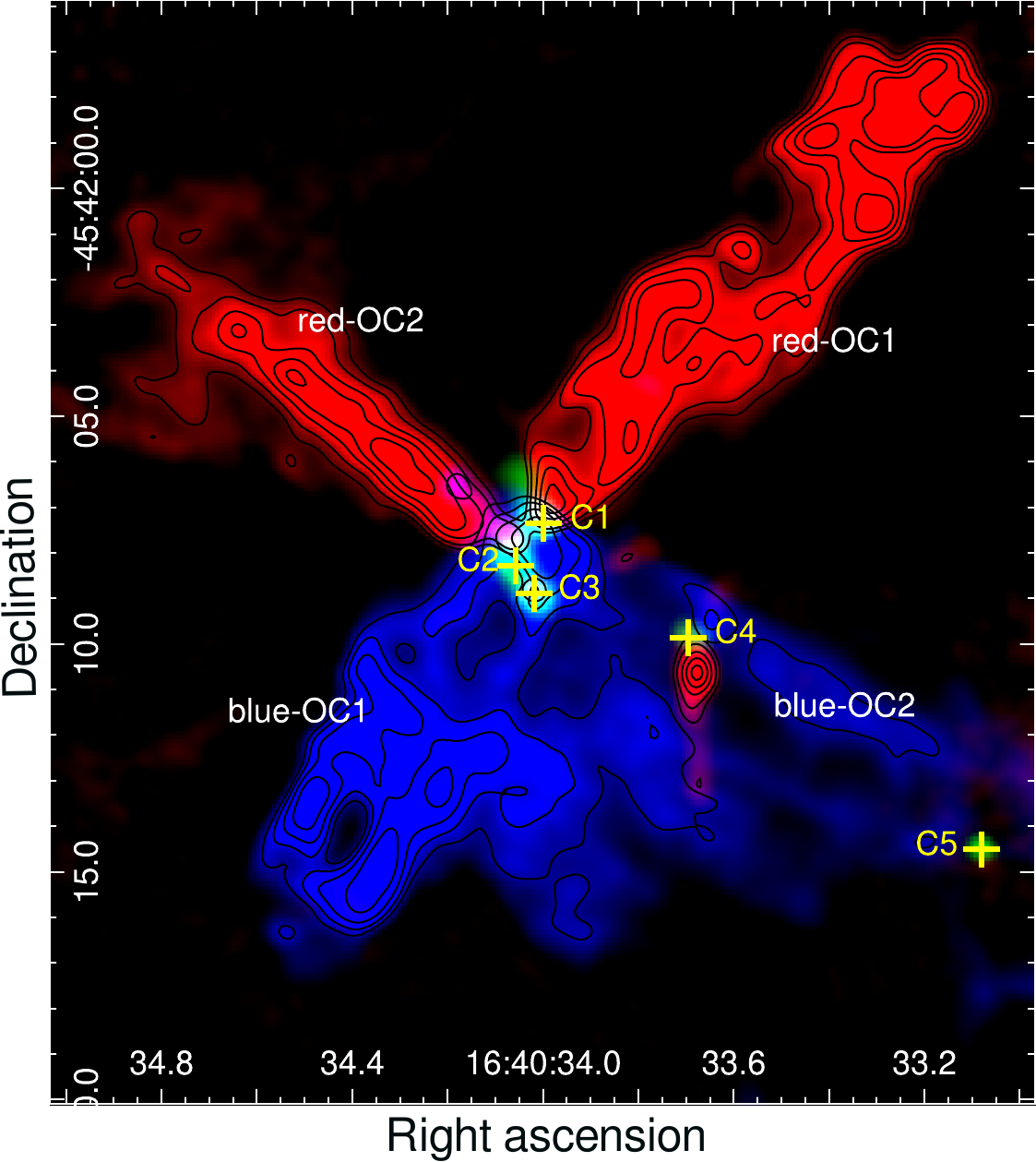}
    \caption{$^{12}$CO J=3--2 emission distribution integrated (moment 0) between $-$90 and $-$70 \ks~(blue), and between $-$55 and $+$5 \ks~(red). The systemic velocity of the complex is about $-$64 \ks. The black contour levels are at 5, 8, 12, and 18~mJy beam$^{-1}$. The ALMA continuum emission at 340 GHz is represented in green. The position of the five molecular cores is indicate with the yellow crosses.  For more details see \citet{ortega2023}.}
    \label{fig:outflows_mom0}
\end{figure}

EGO\,G338 is a star-forming region embedded in the massive molecular clump AGAL G338.9188+0.5494, located at a distance of about 4.4 kpc with a systemic velocity of approximately $-$64.1 \ks~\citep{wienen2015}. 

\citet{ortega2023}, using high-angular resolution data from the Atacama Large Millimeter Array (ALMA), revealed that the ATLASGAL clump is fragmented into at least five cores (C1 to C5), with molecular outflow activity mainly associated with cores C1 and C2 as observed in CO J=3$-$2 emission. While the red lobes, named red-OC1 and red-OC2 of these outflows (see Fig. \ref{fig:outflows_mom0}), are spatially separated and relatively well-collimated, the blue lobes (blue-OC1 and blue-OC2) seem to be mixed, appearing as a single cone-like structure opening toward the south. In \citet{ortega2023}, this morphology was attributed to the scattering of the blue-OC2 molecular outflow by the core C3. However, considering that the outflows are extended structures likely propagating very close to one another, based on the intriguing observed blue-shifted CO feature morphology, we suggest that, in addition to this hypothesis, we should consider a possible scenario of outflow collision. Therefore, the present study aims to analyse this possibility through both observations and numerical hydrodynamical simulations. 

\begin{figure*}
    \centering
    \begin{subfigure}{0.40\textwidth}
        \makebox[\linewidth][l]{%
            \includegraphics[width=\linewidth]{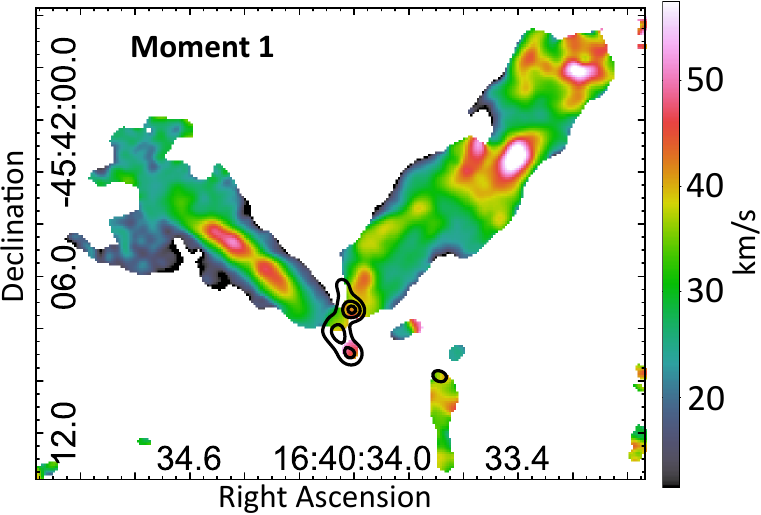}%
        }
    \caption{}
    \end{subfigure}
    \hfill
    \begin{subfigure}{0.545\textwidth}
        \makebox[\linewidth][l]{%
            \hspace*{-0.4cm}
            \includegraphics[width=\linewidth]{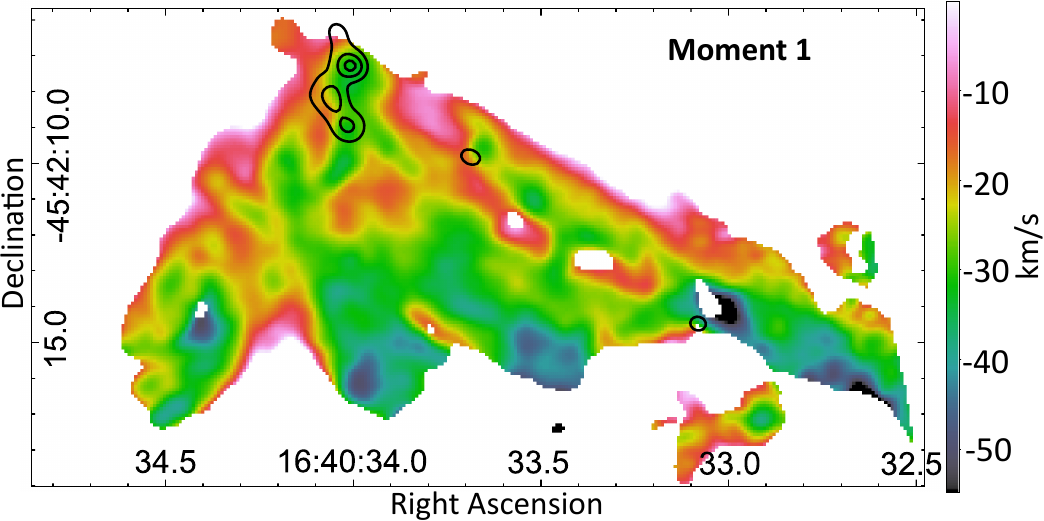}
        }
       \caption{}
    \end{subfigure}
    \vspace{0mm} 
    \begin{subfigure}{0.4\textwidth}
        \makebox[\linewidth][l]{%
            \includegraphics[width=\linewidth]{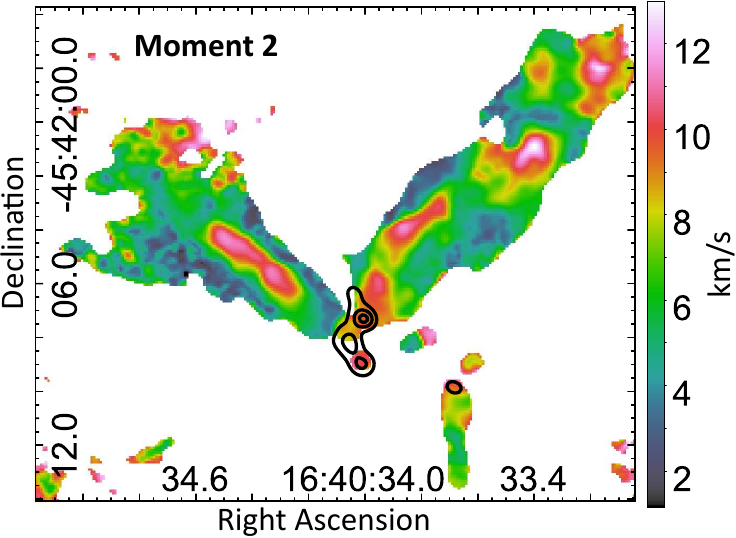}%
        }
    \caption{}
    \end{subfigure}
    \hfill
        \begin{subfigure}{0.54\textwidth}
        \makebox[\linewidth][l]{%
        \hspace*{-0.5cm}
            \includegraphics[width=\linewidth]{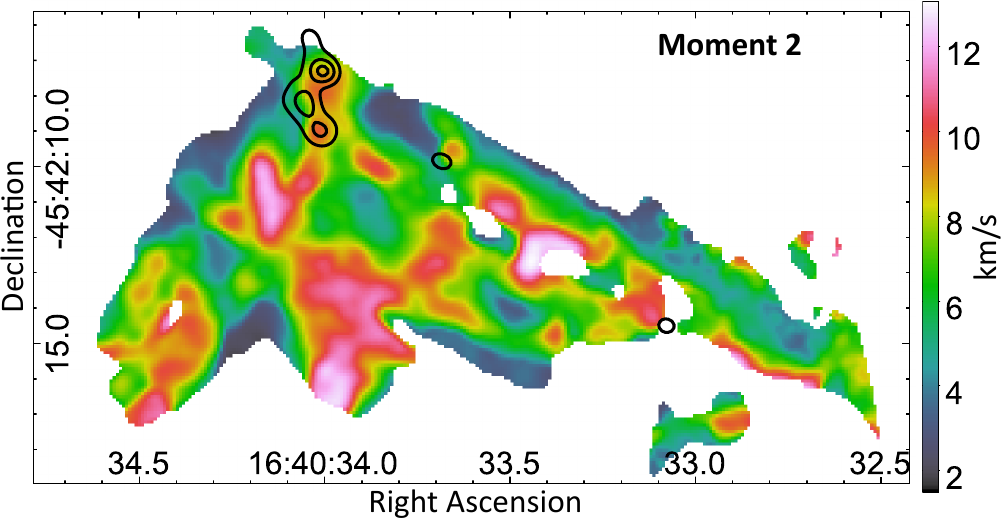}%
        }
    \caption{}    
    \end{subfigure}
    \caption{$^{12}$CO J=3--2 moment 1 (mean velocities) and moment 2 (velocity dispersion) map in colour scale. Black contours show the ALMA continuum emission at $340$ GHz, with levels of $60$, $100$, $200$ mJy beam$^{-1}$. (a) Moment 1 map, between $0$ and $55$ km s$^{-1}$. (b) Moment 1 map, between $0$ and $-55$ km s$^{-1}$. In the case of moment 1 maps, for simplicity, the systemic velocity of the region was set at v$_{\rm LSR}=0$ \ks. (c) Moment 2 map, between $0$ and $55$ km s$^{-1}$. (d) Moment 2 map, between $0$ and $-55$ km s$^{-1}$.}
    \label{fig:moments}
\end{figure*}

\section{ALMA observations: main results}
\label{ALMA_observations}

The used ALMA data are presented in detail in \citet{ortega2023}. Data cubes from project 2017.1.00914 (PI: Csengeri, T.; Band 7) were obtained from the ALMA Science Archive\footnote{http://almascience.eso.org/aq/}. 
The angular and spectral resolutions are 0\farcs47 and 1.1~MHz, respectively. The velocity resolution is about 1.0~kms$^{-1}$. The rms noise level is 3.6~mJy~beam$^{-1}$ for the emission line (averaged each 10~km s$^{-1}$) and 0.2~mJy~beam$^{-1}$ for the continuum emission. The beam size of the 340~GHz continuum data provides a spatial resolution of about 0.01~pc ($\sim$ 2000 au) at the distance of 4.4~kpc, which is appropriate for spatially resolving the substructure of the molecular outflows related to EGO G338.

As presented in \citet{ortega2023} and mentioned above, in Fig.\,\ref{fig:outflows_mom0} we show a three-colour composite image of the EGO G338 region. Yellow crosses indicate the positions of the five identified molecular cores. The continuum emission at 340 GHz is presented in green. The integrated $^{12}$CO J=3--2 emission between $-$55 and +5 \ks, displayed in red, corresponds to the red-shifted lobes of the molecular outflows originating from the cores C1 and C2. Finally, in blue, it is displayed the integrated $^{12}$CO J=3--2 emission between $-$90 and $-$70\ks, showing the merged blue-shifted lobes of the molecular outflows associated with the C1 and C2 cores extending southwards.

A striking feature is the difference in the morphologies of the northern and southern outflows. While the northern red-shifted lobes appear to be well collimated and remain spatially separated, the southern blue-shifted lobes appear to be merged into a single conical structure. In particular, the blue-OC1 lobe appears somewhat shorter than the red-OC1 lobe and shows some evidence of deflection.  
By reanalysing the blue-shifted southern structure morphology and considering both the direction of the blue lobes and the low plausibility that an outflow-core collision can generate such a spatially very extended structure, we suggest an alternative interpretation as presented in \citet{ortega2023}. We wonder if the conical structure resulting from merging gas belonging to the blue lobes is caused by a collision between outflows.

To evaluate the gas kinematics of the observed molecular outflows, in Fig.\,\ref{fig:moments} we present maps of moments 1 and 2 of the CO emission. In the case of moment 1 maps, for simplicity, the systemic velocity of the region was set at v$_{\rm LSR}=0$ \ks. The black contours represent the 340 GHz ALMA continuum emission at the 10$\sigma$ rms level, indicating the position of the molecular cores. 

From the moment maps, it can be noticed that the red-shifted lobes of the molecular outflows, the northern ones, show quite ordered layer structures, with maximum velocity and velocity dispersion predominantly localised at inner layers (see Fig.\,\ref{fig:moments}-(a) and (c)). In the case of the blue-shifted counterpart components, the southern ones, they exhibit a more complex structure with regions of high-velocity and high-velocity dispersion located mainly at the central part of the cone-like feature  (see Fig.\,\ref{fig:moments}-(b) and (d)). This shows the presence of an extended, uncollimated feature of turbulent gas south of the region where the outflows are generated. Without discarding some probable contribution of a possible interaction between the outflows and the quiet cores, in what follows, using numerical simulations, we evaluate whether a collision between outflows can produce this kind of extended turbulent gaseous structure. 

\section{Numerical simulations}

The numerical study was performed using 
the parallel 3D HD code {\sc guacho} \citep{Esquivel2009, Villarreal2018}. This code solves the ideal gas dynamical equations on a fixed Cartesian grid, reading as follows:

\begin{equation}
    \frac{\partial\rho}{\partial t}+\nabla\cdot(\rho\mathbfit{u})=0\;,
	\label{eq:mass}
\end{equation}

\begin{equation}
    \frac{\partial(\rho\mathbfit{u})}{\partial t}+\nabla\cdot\left[\rho\mathbfit{u}\otimes\mathbfit{u}+\mathbb{I} p\right]=0\;,
	\label{eq:momentum}
\end{equation}

\begin{equation}
    \frac{\partial e}{\partial t}+\nabla\cdot \left[
    \left(e + p\right)\mathbfit{u}
    \right]=Q_L\;,
	\label{eq:energy}
\end{equation}
In the equations, $\rho$ represents the mass density, $\mathbfit{u}$ the velocity vector, $p$ the gas pressure, and $e$ indicates the total energy density. 
In Eq.~(\ref{eq:momentum}), $\mathbb{I}$ denotes the identity matrix. The energy density is given by $e=\rho u^2/2+p/(\gamma-1)$, 
where $\gamma$ is the heat capacity ratio of the gas, set to 5/3. In addition to solving the gas dynamical equations, the code integrates an extra continuity equation for a passive scalar. 
This scalar is used to track different gas components (e.g., jet material, circumstellar gas).

The code also incorporates radiative cooling, defined as $Q_L=n^2 \Lambda(T)$ (see Eq.~\ref{eq:energy}), where $n$ is the gas number density and $\Lambda(T)$ is a parameterised function of the temperature 
that describes optically-thin cooling \citep{Dalgarno1972}.
For numerical integration of Eqs.~(\ref{eq:mass})--(\ref{eq:energy}) over time, the code employs a second-order Godunov method with the approximate Riemann solver HLLC \citep{Toro1994}. Zero-gradient (outflow) boundary conditions are applied at the edges of the computational domain.

\subsection{Initial numerical setup}
\begin{figure*}
    \centering
    \includegraphics[width=16cm]{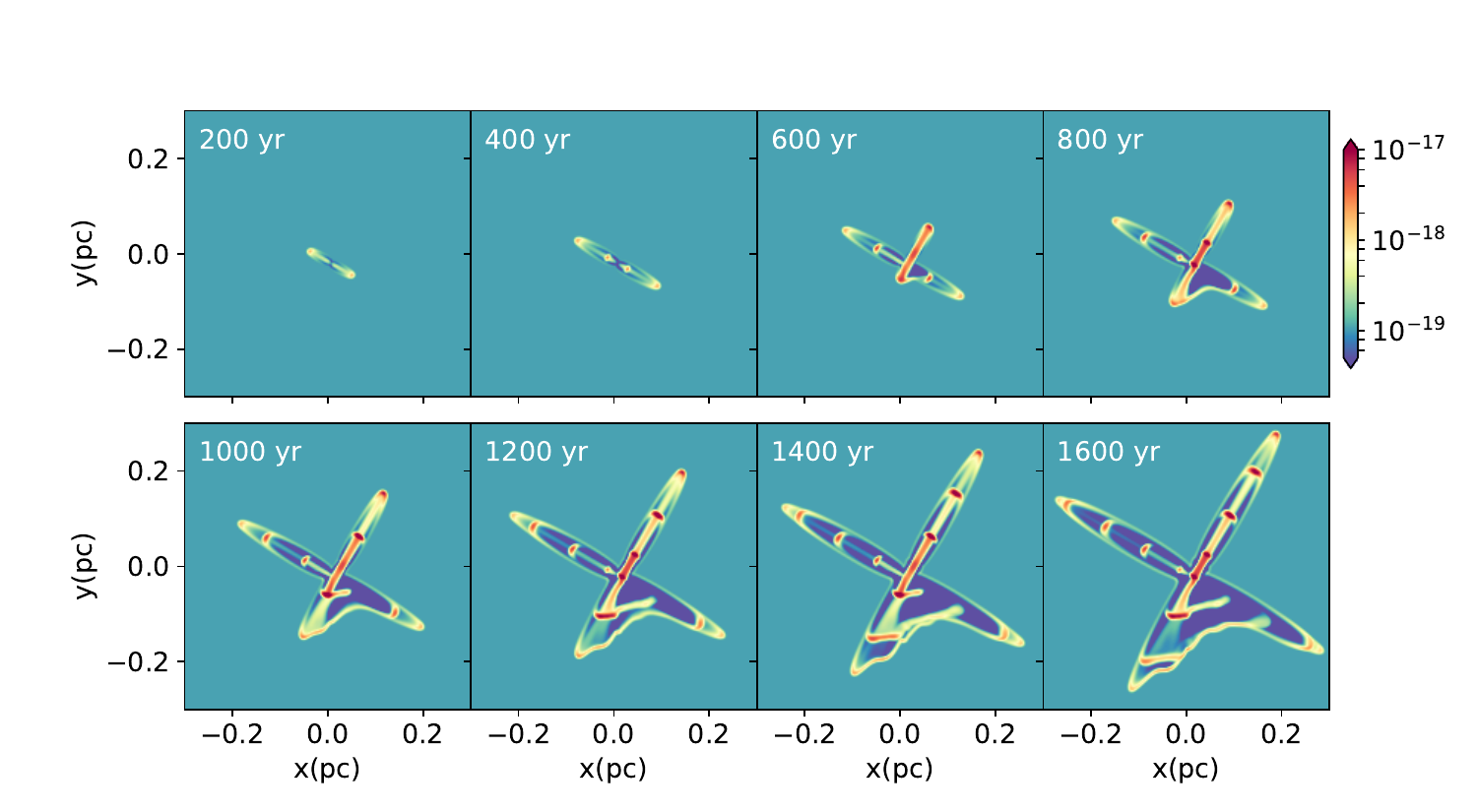}
    \caption{Temporal evolution of the density distribution on the $y'=0$ plane. The logarithmic colour bar gives the density in units of g cm$^{-3}$. Both axes are given in parsecs.}
    \label{fig:devol}
\end{figure*}

\begin{figure*}
    \centering
    \includegraphics[width=14cm]{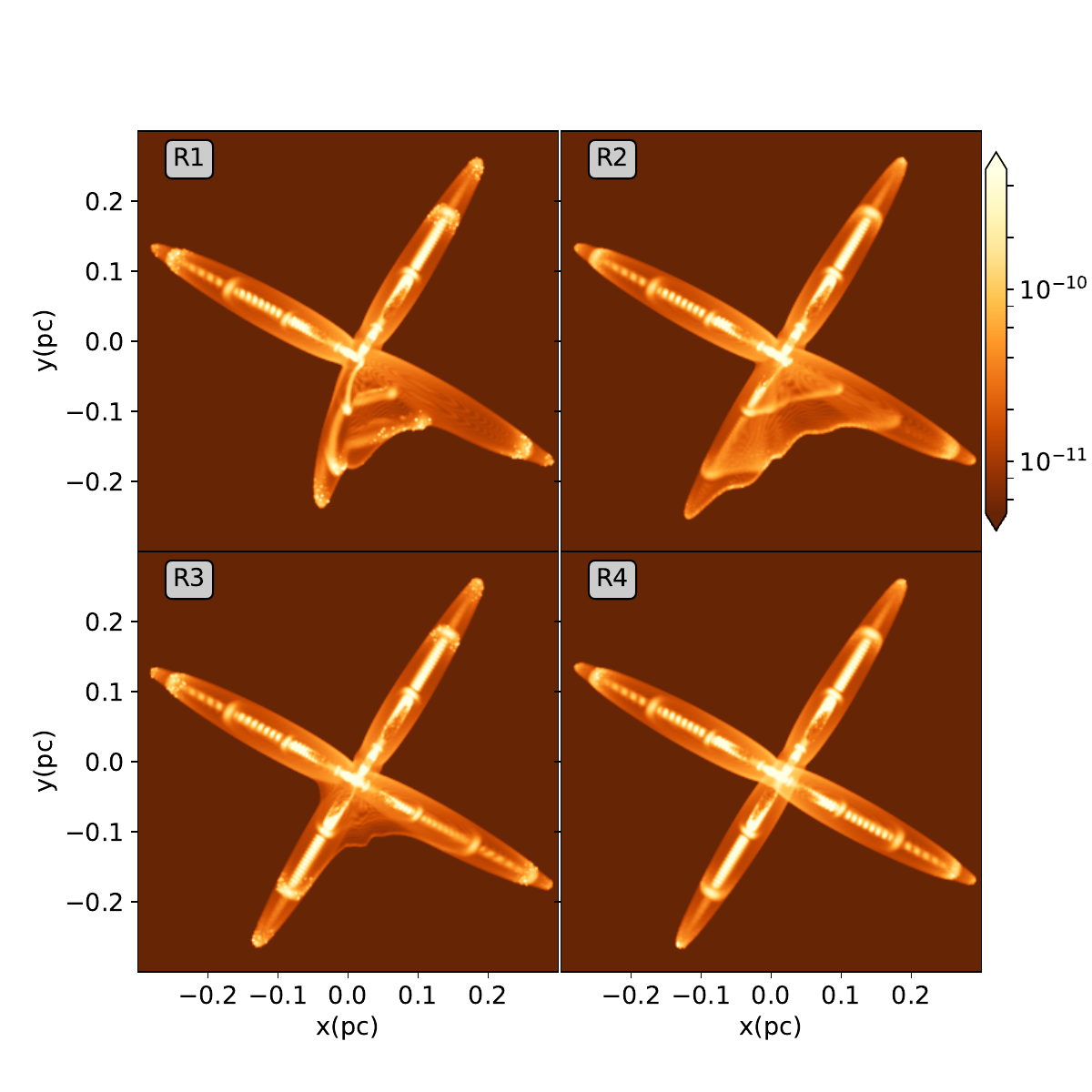}
    \caption{Comparison of the $^{12}$CO J=3--2 emission obtained at $t=1600$~yr from all runs. The logarithmic colour bar represents the emission in units of erg cm$^{-2}$ s$^{-1}$ sr$^{-1}$}
    \label{fig:compICO}
\end{figure*}

\begin{figure*}
    \centering
    \includegraphics[width=14cm]{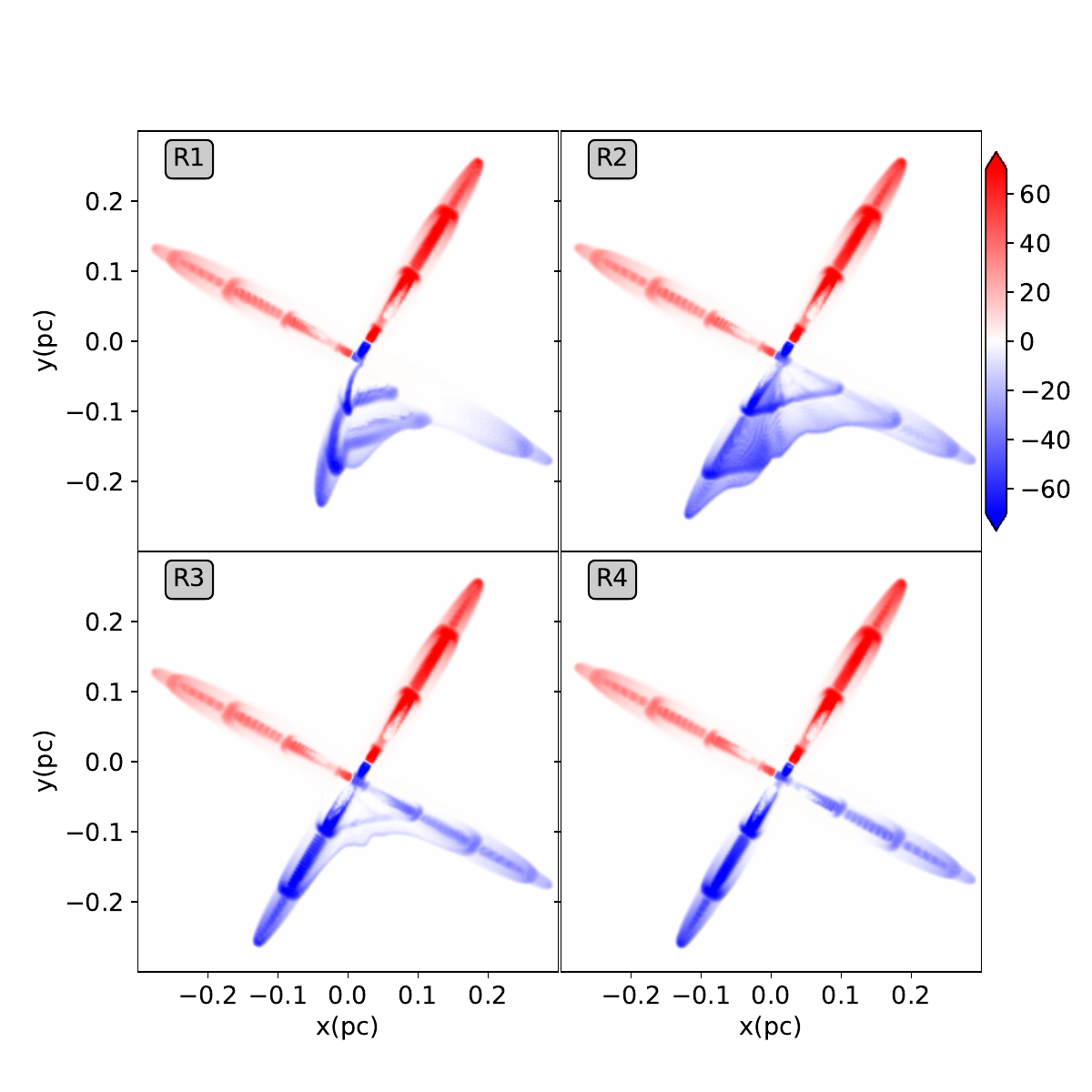}
    \caption{Comparison of the first moment ($\bar{v}$) obtained from all runs. The linear colour bar is the $\bar{v}$ in km s$^{-1}$. Both axes are in pc.}
    \label{fig:compmom1}
\end{figure*}

\begin{figure*}
    \centering
    \includegraphics[width=14cm]{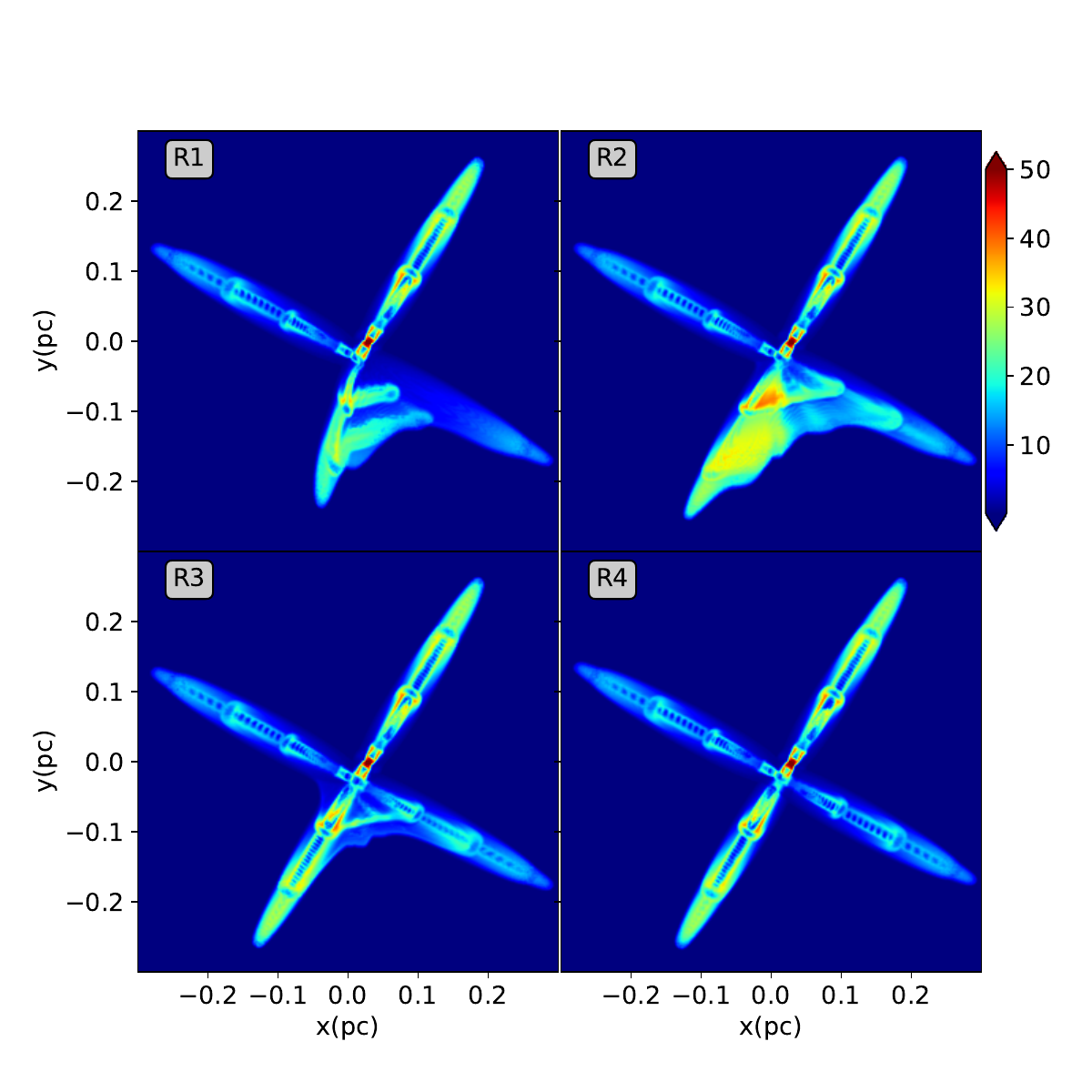}
    \caption{Comparison of the dispersion velocity ($\sigma$) maps obtained at $t=1600$~yr from all runs. The colour bar gives $\sigma$ in km s$^{-1}$}
    \label{fig:compmom2}
\end{figure*}

Numerical simulations were carried out in a 3D Cartesian cubic domain with dimensions of 0.6~pc on each side and a spatial resolution of $10^{-3}$~pc. Within this computational domain, we established an environment characterised by a uniform number density of $10^4$~cm$^{-3}$ and a temperature of 10~K.

The goal of this study is to reproduce the morphology and kinematics of the outflows observed in EGO G338. To achieve this, we performed several simulations that consider scenarios in which interactions between the outflows of EGO G338, specifically OC1 and OC2, are likely to occur, following the ideas given by \citet{cunningham2006}.

We defined two cylindrical jets with radii  \(r_j=5 \times 10^{-3}\)~pc and lengths \(l_j=10^{-2}\)~pc, which inject mass into the surrounding medium at a rate $\dot{\mathrm{M}}_j$. Additionally, we implement time-dependent ejection velocities for jets, given by:
\begin{equation}
v_j=v_0 (1+\Delta v_j \cos{(2\pi/\tau_j t)})
\label{eq:vjet}
\end{equation}
where $v_0$ is the mean jet velocity, $\Delta v_j$ is the velocity variation, and $\tau_j$ is period. For both jets, we set $\Delta v=0.5$ and $\tau_j=400$~yr. In Table \ref{tab:jetpar} we list the jet parameters used for the case of OC1 and OC2 outflows. The angles $\theta$ and $\phi$ represent the angular positions of the jet axis relative to the $\hat{z}$ and the $\hat{x}$ directions, respectively.

We have two reference systems: the $x'y'z'$ system, which corresponds to the computational domain, and the $xyz$ system, which represents the observational framework. In this observational system, the $xy$ plane represents the plane of the sky, and the $z$ axis corresponds to the line of sight. Initially, these two systems align perfectly. To generate the synthetic maps, we rotate the $x'y'z'$ system relative to the observational system to achieve a better fit with the observations.

The central positions of the cylinders used in our simulations are outlined in Table \ref{tab:jetpos}. After rotating the computational domain ($x'y'z'$ system) by $20^\circ$ relative to the plane of the sky, the jet centres align with the positions of the OC1 and OC2 sources as reported by \citet{ortega2023}.

\begin{table}
\caption{Jet parameters}
	\centering
        \begin{tabular}{ccccrr} 
		\hline
        \hline
		Jet &  $\dot{M}_{\text{j}}$ (${\textrm M}_\odot\ \textrm {yr}^{-1}$) & $v_0$ (km s$^{-1}$) & $\theta_\textrm{j}$ ($^{\circ}$) & 
        $\phi_\textrm{j}$ ($^{\circ}$) &
    $\tau_{\textrm{on}}$ (yr)\\
		\hline
		OC1 & $1.3\times 10^{-3}$ & $225.$ & 60. & 0. & 400.\\
		OC2 & $6.5\times 10^{-4}$ & $200.$ & 30. & 180. & 0.\\
		\hline
	\end{tabular}

	\label{tab:jetpar}
\end{table}

\begin{table}
\caption{Jet source positions}
\centering
\begin{tabular}{cclllr}
\hline
\hline
Run & jet & $x_0(pc)$ & $y_0(pc)$ & $z_0(pc)$ & $b(r_j)$ \\
\hline
R1& OC1 & 0.03 & 0. & 0.& 0. \\
 & OC2 &0.007 & 0. & -0.019 & \\
 \hline
R2 & OC1 & 0.03 & 0. & 0. & 1. \\
 & OC2 & 0.007 & -0.005 & -0.02 & \\
 \hline
R3 & OC1 & 0.03 & 0. & 0. & 2. \\
 & OC2 & 0.007 & -0.01 & -0.028 & \\
 \hline
R4 & OC1 & 0.03 & 0.05 & 0.017 & 20. \\
 & OC2 & 0.007 & -0.05 & -0.035 & \\
 \hline
\end{tabular}

\label{tab:jetpos}
\end{table}

\subsection{Synthetic CO emission}
To facilitate comparison with observational data, we constructed Position-Position-Velocity (PPV) cubes of the $^{12}$CO J=3--2 transition from the $n$, $T$, and $v$ distributions derived from our numerical results.

The emissivity of the $^{12}$CO J=3--2 transition is determined by employing the Boltzmann excitation equation along with the partition function \citep{anglada1999,nhung2017}. We can estimate it as follows:
\begin{equation}
    \epsilon_{32}(x,y,z,v_r) = \frac{h \nu_{32}}{4\pi} \frac{1}{Z(T)} g_{i}\exp\bigg(-\frac{h \nu_{32}}{k_B T}\bigg)
    \chi_{co} n A_{32} \phi(v_r),
    \label{eq:emisco}
\end{equation}
where $\chi_{co}=1.67 \times 10^{-4}$ is the fractional abundance of the CO concerning the total gas density $n$, $Z(T)=T /\Theta_{\textrm{rot}}$ (being $\Theta_{\textrm{rot}}=2.8$~K the rotational temperature), $h$ is the Planck's constant, $k_B$ is the Boltzman's constant, $\nu_{32}=345$~GHz is the transition frequency, $A_{32}= 2.5\times 10^{-6}$~s$^{-1}$  is the Einstein coefficient, and $g_i =2i+1=7$ (with $i=3$) is the statistical weight.
The Gaussian line profile $\phi(v)$ is given by:
\begin{equation}
    \phi(v_r)=\sqrt{\frac{m_{co}}{2\pi k_B T}}\exp\bigg(-\frac{m_{co}(v_r-v)^2}{2 k_B T} \bigg),
\end{equation}
being $v_r$ the channel velocity, $v$ the gas velocity along the line of sight, and $m_{co} =3.32\times 10^{-23}$~g the mean molecular mass.

We obtain the PPV cubes integrating the $\epsilon_{32}(v)$ given by Eq.(\ref{eq:emisco}) along the line of sight as:
\begin{equation}
    \textrm{PPV}(x,y,v_r)=\int_{\textrm{LoS}} \epsilon_{32}(x,y,z,v) dz
    \label{eq:ppv}
\end{equation}

Once we obtain the PPV cubes, we can create emission CO maps by adding the contributions from all the velocity channels as:
\begin{equation}
I(x,y)=\int_{LoS} \mathrm{PPV}(x,y,v_r) dv_r
\label{eq:Ico}
\end{equation}

To better understand the flow kinematics, we can generate first-moment (centroid) maps from the PPV cube by means of:
\begin{equation}
    \bar{v}(x,y)=\frac{\int v_r \mathrm{PPV}(x,y,v_r) dv_r}{I(x,y)}
    \label{eq:vmed}
\end{equation}
These maps will provide valuable insights into the mean velocity at each position in the $xy$ plane (the plane of the sky).

Furthermore, constructing second-moment maps from the PPV cubes will enable us to assess the level of turbulence and velocity dispersion, enhancing our overall analysis. We get the velocity dispersion maps by:
\begin{equation}
    \sigma(x,y)=\sqrt{\frac{\int (v_r-\bar{v}(x,y))^2 \mathrm{PPV}(x,y,v_r) dv_r}{I(x,y)}}
    \label{eq:sigmav}
\end{equation}

\subsection{Numerical results}

As discussed in Section 4.1, we conducted an investigation into the evolution of two cylindrical jets separated by varying impact parameters. The objective was to identify specific observational characteristics that may indicate interaction between these jets. To facilitate the analysis of the numerical results, we will use the same names assigned to the outflows in the observations as those given to their similar counterparts in the simulations.

Our preliminary analysis focused on the temporal changes in the density distribution.

For example, Figure \ref{fig:devol} illustrates the time evolution of run R2, where \( b_j = r_j \). Through our initial tests, we determined that to achieve the observed dimensions of the EGO G338 system, the simulated jets required igniting the OC2 jet before the OC1 jet. Accordingly, we opted to establish a 400-yr delay between the two ignitions. 

This displays the evolution of the density distribution within the central plane of the computational domain. For up to 400~yr, the red and blue OC2 outflows have developed symmetrically. After 600~yr of evolution, the interaction between the blue flows of OC1 and OC2 becomes evident. As time progresses, the blue flow of OC2 becomes increasingly different from its red counterpart, leading to the formation of a more hollow cavity. Additionally, a conical region appears between the two blue outflows. At 1.6 kyr, this region is characterised by a series of dense filaments.

Figure \ref{fig:compICO} compares the CO emission maps generated for all simulations at an evolution time of 1.6~kyr. In the case of run R1 (with zero impact parameter), the most significant observation is that the blue OC1 outflow exhibits a pronounced deviation in its trajectory. This alteration occurs due to its collision with the blue OC2 outflow, as illustrated in the top-left panel of this figure. Besides, while the northward emission clearly indicates two outflows, an extensive region is noted between the southerly outflows.

In the simulation map for run R2, where the impact parameter $b_j$ equals the jet radius, the blue OC1 outflow shows less deflection compared to run R1. Regardless, the area between the two outflows in the southern region is larger, providing further evidence of the interaction between the OC1 and OC2 flows.

The interaction features observed in runs R1 and R2 are noticeably less prominent in the emission map of run R3, where the impact parameter is equivalent to 2~$r_j$.

The R4 map outlines a scenario involving two non-interacting outflows aligned along the line of sight. Both pairs of red and blue outflows are distinctly observable, with no emissions evident in the space between the blue ones. The synthetic maps produced for this run indicate a slight increase in CO emissions in the region where the outflows appear to intersect. This observed increase is attributed to the addition of the CO emission from these outflows along the line of sight, resulting in a projection effect. 

Figure \ref{fig:compmom1} compares the mean velocity or moment 1 maps generated by all models. The values and distribution obtained for R1 and R2 are in good agreement with those observed in the upper maps of Figure \ref{fig:moments}.

A noteworthy finding has emerged from the analysis of the second-moment maps. Figure \ref{fig:compmom2} clearly indicates that when the impact parameter is set to \( r_j \), there is a notable increase in velocity dispersion. Conversely, when the impact parameter is specified as zero, the observed increase in dispersion is less pronounced. This phenomenon is similarly evident when the distance between the jets is \( 2r_j \).

The numerical results suggest that the observed OC1 and OC2 outflows are indeed interacting. The scenario that most effectively explains the observed CO emission and the dispersion velocity distribution of the EGO G338 region is represented by the run R2, i.e. when the impact parameter is $r_j$ (upper-left panels of Figs. \ref{fig:compICO} and \ref{fig:compmom2}). On the other hand, during our analysis of run R4, we observed that the blue and red outflows related to OC1 and OC2 display symmetrical patterns in the emission, moment 1, and moment 2 synthetic maps because there is no interaction between the two outflows (see lower-left panel of Figs. \ref{fig:compICO}, \ref{fig:compmom1}, and \ref{fig:compmom2}).

\section{Discussion}

In this section, a comparison is presented between the main results of observations and numerical simulations, along with an estimate of the collision probability of molecular outflows for the observed geometric configuration.

\subsection{Comparison between observations and numerical simulations}

Numerical simulations were performed using input parameters consistent with the observational data presented by \citet{ortega2023} and reanalysed in Sect.\,\ref{ALMA_observations}. However, it is important to note that the dynamical timescales derived in the former observational study were not used as inputs to determine which core initiated the outflow activity first. While the dynamical times for the outflows related to cores C1 and C2, estimated to be 4.2 and 2.7 kyr, respectively, suggest that core C1 initiated the molecular outflow activity earlier, the numerical results indicate otherwise. Our simulations suggest that the observed southern gaseous conical structure appears, and the blue-OC2 lobe reaches its observed spatial extension only when core C2 is the first outflow launched. This issue highlights the inherent uncertainty of dynamical timescales estimated from observational data, suggesting that one must be cautious when making such estimations, especially in a likely context of outflow collision.

Figure\,\ref{fig:compICO} shows that, among all runs, R2 ($b_j=r_j$) provides the best match to the observations, exhibiting a smaller deflection of the structure associated with the blue-OC1 lobe and a broader spatial extent of the conical structure, consistent with the observations. Furthermore, the moment 2 map of the R2 run reveals an increase in velocity dispersion in the inter-blue lobe region (Fig. \ref{fig:compmom2}), closer to the blue-OC1 lobe structure, which is also evident in the moment 2 map obtained from the ALMA observations (Fig. \ref{fig:moments}).

\subsection{Outflow-outflow interaction probability}

The mass ejection of low-mass protostars has been observed to have a preferential direction because these outflows appear to be collimated by the mechanism that ejects them. These outflows do not remain collimated throughout their entire evolution; however, they widen due to their interaction with the surrounding environment. It is because the deceleration of the injected flow, or jet, decreases the magnitude of the hydrodynamic pressure until it is on the same order as the thermal pressure. At that point, the lateral expansion of the outflows begins to form a cavity that moves primarily parallel to the jet but grows slightly in the transverse direction, dragging along the material from the surrounding medium that is in an envelope which encloses the entire outflow. This type of outflow, in three-dimensional space, has a probability of encountering another one produced by a nearby object; however, it has been observed that these objects have an opening that depends on the distance along the outflow. Therefore, the probability of this type of object encountering another in space is very low. In \citet{Raga2025}, the likelihood of such interactions occurring was estimated, and the probability was related to the separation distance between two or more injection sources. They also considered the opening angle of the gas injection in relation to the probability of collision between two or more different jets. In \citet{RG25}, they studied the effect of the interaction of two flows ejected by low-mass protostars, considering not only the main flow due to the jets, but also the interaction between the envelopes—that is, the molecular flow produced by the injection of material from two sources. It, of course, increases the probability of interaction because the envelopes are structures that occupy a larger volume in space.

\begin{figure}
    \centering
    \includegraphics[width=1\columnwidth]{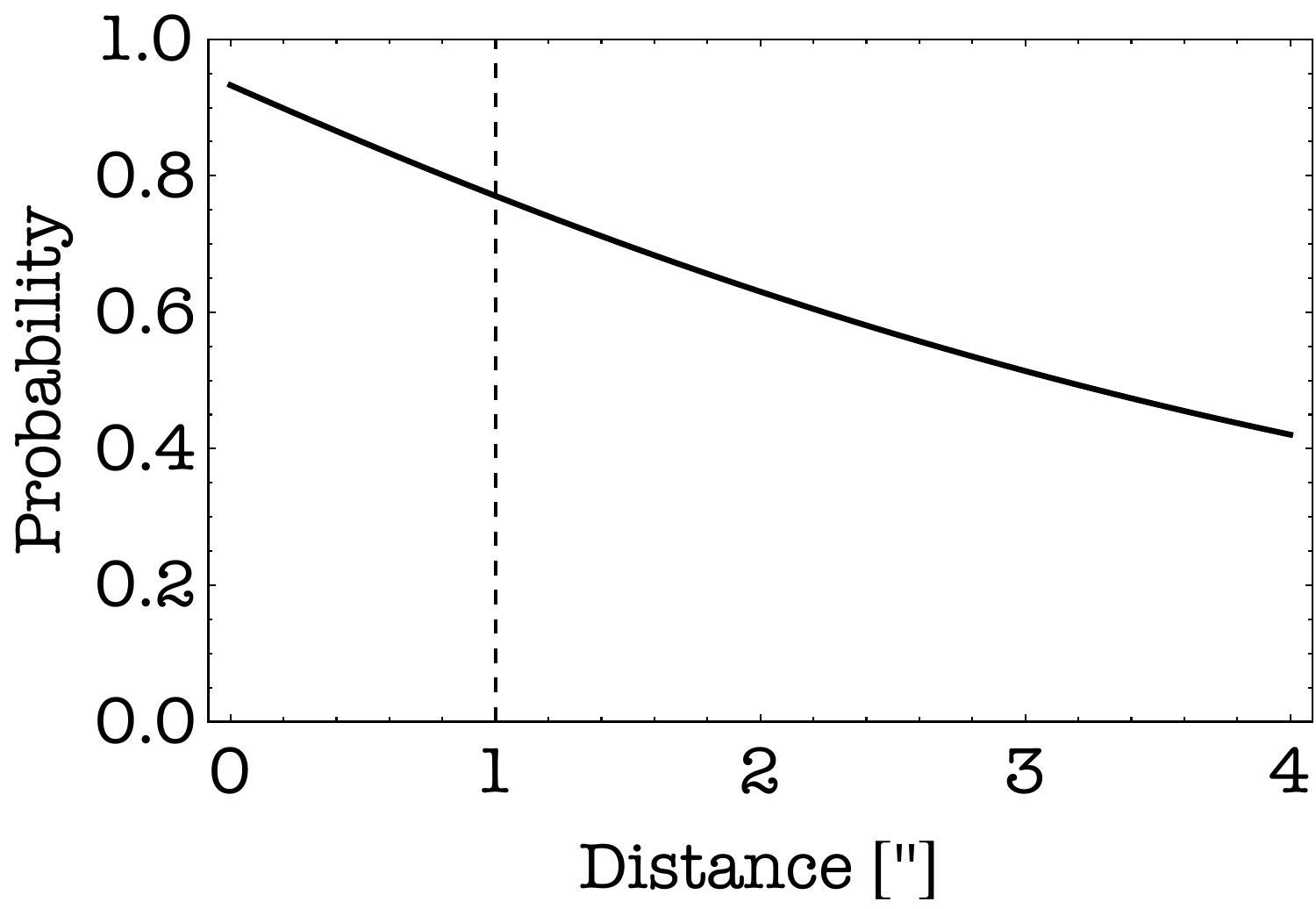}
    \caption{Probability of encounter between the outflows produced by C1 and C2 as function of their distance (Continous black line). The dotted line represents the observed projected distance between the sources.}
    \label{fig:prob}
\end{figure}

Here, we present an alternative method, also applied in \citet{RG25}, to quantify the probability of two outflows interacting in EGO G338, ejected from two protostellar sources that are a distance $r$ apart. First, we can simplify the morphology of an outflow as a bicone with semilength $L$ and semiaperture angle $\alpha$, then we assume that an observer located in one of the sources can see the outflow produced by the other source projected as two spherical triangles with a shared apex from their point of view. Then, the probability of an encounter of both outflows is the ratio between the solid angle $\Omega+\omega$ occupied by both outflows and the solid angle of half a sphere

\begin{equation}
    P=\frac{\Omega+\omega}{2\pi},
    \label{eq:probability}
\end{equation}
where $\Omega=2\phi(1-\cos\theta_b)$ is the solid angle of the observed outflow, $\tan\theta_b=L/r$, $\phi=2\sin\theta_b\tan\alpha$ and $\omega=2\pi(1-\cos\beta)$, assuming $\beta$ is the aperture semiangle of the outflow produced by the observer. This analysis does not take into account border effects, that is, if $\Omega+\omega>2\pi$, then $P=1$. Figure \ref{fig:prob} shows probability with $r$ as a free parameter, since every other parameter can be obtained directly from the observation (Figure \ref{fig:outflows_mom0}), where $\alpha=40^\circ$, $L=8''$, and $\beta=25^\circ$.  Even when the outflows are larger that $4''$, at this angular distance, the bicone approximation holds.  $P$ decays as the sources are farther away (solid line) and the maximum probability is given by the observed projected separation of C1 and C2, (dotted line), where $P\sim 80\%$.

\section{Conclusions}

We present a novel interpretation of outflows in the EGO G338 region, with a specific focus on the potential interaction between  OC1 and OC2 outflows. Some studies in the literature, including those conducted by \citet{zapata2018}, have examined comparable scenarios. The main objective of our investigation was to evaluate the probability that such an interaction occurs within the EGO G338 region. Using observations of this astrophysical object by \citet{ortega2023} and prior research by \citet{Raga2025,RG25,zapata2018,cunningham2006}, we calculated a collision probability of 80\%. This significant probability provides substantial support for the proposed interaction scenario.

Henceforth, we conducted a study employing three-dimensional hydrodynamic simulations to explore the likely collision scenarios involving the two jets. The primary objective of this analysis was to improve our understanding of the observed morphology and kinematics within the EGO G338 region. From our numerical simulations, we generated synthetic PPV cubes of the $^{12}$CO J=3-2 emission to compare directly with the observational data.

From the analysis of these PPV cubes, we developed synthetic maps showing CO emission, mean velocity, and velocity dispersion along the line of sight. Our comparative analysis indicates that the observed morphology—characterised by extensive emission associated with southward outflows—and the kinematics, particularly the high velocity dispersion within that region, are consistent with a scenario in which the blue lobes of the OC1 and OC2 outflows 
are interacting at an impact parameter equivalent to $r_j$. This result provides new evidence for the existence of a phenomenon not so commonly observed in star formation: the collision of molecular outflows associated with nearby protostars.




\section*{Acknowledgements}

PFV, ARG, EAP, and AE acknowledge financial support from PAPIIT (DGAPA, UNAM) grant AG101125. PFV thanks the fellowship from the PASPA program (DGAPA, UNAM). MO and SP are members of the Carrera del Investigador Científico (CONICET, Argentina). This work was partially supported by the Argentinian grants PIP 2021 11220200100012 and PICT 2021-GRF-TII-00061 awarded by CONICET and ANPCYT. We thank Enrique Palacios (ICN-UNAM) for maintaining the cluster where the simulations presented in this work were performed. EAP acknowledges scholarship
from SECIHTI-México 4050143. PFV dedicates this work to the memory of his colleague and friend, Dr Daniel Osvaldo Gómez (IAFE, UBA).

\section*{Data Availability}


The data underlying this article will be shared on reasonable request to the corresponding author.
The observational data underlying this paper were accessed from the ALMA Science Archive at  http://almascience.eso.org/aq/



\bibliographystyle{mnras}
\bibliography{example} 








\bsp	
\label{lastpage}
\end{document}